\documentclass{aa}  

\usepackage{txfonts}
\usepackage{natbib,twoopt}
\usepackage[breaklinks=true]{hyperref} 
\bibpunct{(}{)}{;}{a}{}{,}             
\makeatletter
  \newcommandtwoopt{\citealtads}[3][][]{\href{http://adsabs.harvard.edu/abs/#3}%
    {\def\hyper@linkstart##1##2{}%
     \let\hyper@linkend\@empty\citealp[#1][#2]{#3}}}
  \newcommandtwoopt{\citeads}[3][][]{\href{http://adsabs.harvard.edu/abs/#3}%
    {\def\hyper@linkstart##1##2{}%
     \let\hyper@linkend\@empty\citealp[#1][#2]{#3}}}
  \newcommandtwoopt{\citepads}[3][][]{\href{http://adsabs.harvard.edu/abs/#3}%
    {\def\hyper@linkstart##1##2{}%
     \let\hyper@linkend\@empty\citep[#1][#2]{#3}}}
  \newcommandtwoopt{\citetads}[3][][]{\href{http://adsabs.harvard.edu/abs/#3}%
    {\def\hyper@linkstart##1##2{}%
     \let\hyper@linkend\@empty\citet[#1][#2]{#3}}}
  \newcommandtwoopt{\citeyearads}[3][][]%
    {\href{http://adsabs.harvard.edu/abs/#3}
    {\def\hyper@linkstart##1##2{}%
     \let\hyper@linkend\@empty\citeyear[#1][#2]{#3}}}
\makeatother
\newcommand{\degree}{^{\circ}}
\newcommand{\dif}{d}
\newcommand{\unit}{}
\begin{document}

\title{Opacity effect on core-shift and the spectral properties of jets}


   \author{R. Sharma 
          \inst{1}
   M. Massi
          \inst{1}
          \and
          G. Torricelli-Ciamponi
          \inst{2}
          }

   \institute{Max-Planck-Institut f\"ur Radioastronomie, Auf dem H\"ugel 69, D-53121 Bonn, Germany\\
              \email{richa.sharma110393@gmail.com}      
         \and
            INAF - Osservatorio Astrofisico di Arcetri, L.go E. Fermi 5, Firenze, Italy}

   \date{Received August 5, 2021; accepted January 27, 2022}

 
  \abstract
{There is theoretical and observational evidence that the jet core position   
changes with frequency. However, the core position for a given frequency may vary with time
in the case of  flares or for a precessing jet.}
 {In this work, we want to explore the changes in core position as a function of
frequency, magnetic field alignment, relativistic electron density, and jet inclination angle.}
 {We use a physical model of a synchrotron-emitting jet. Two cases of the jet are analysed, namely with 
 magnetic field parallel and perpendicular to the jet axis. The evolution of the related spectrum 
 is monitored over the  radio band.}
 {We find  that a smaller jet inclination angle or a higher electron density causes the jet core 
 position to move downstream of the jet and we demonstrate that this displacement of the core 
along the jet gives rise to  a spectral flattening.}
   {}


   \maketitle
%

\section{Introduction} \label{sec:intro}

Relativistic jets are produced when the inflowing, accreting plasma around a  black hole gets expelled and collimated by magnetic forces (e.g. \citealtads{Meier2001}).
When a jet is mapped at high resolution with an interferometer at frequency $\nu$, the surface of the jet where optical depth $\tau_{\nu}\approx 1$ corresponds to the peak flux in the map, and is referred to as the core \citepads{BlandfordKonig1979}. The position of the core is a function of frequency, and this is known as the core-shift effect. 
This characteristic feature of the jet is supported by very long baseline interferometry (VLBI) observations of X-ray binaries (e.g. \citealtads{Paragi1999, Paragi2013, massi2012, Rushton2012}) and active galactic nuclei (AGNs) (e.g. \citealtads{Lobanov1998, Sullivan2009, Hada2011, Sokolovsky2011}).

The core-shift effect is essential for both astrophysical and astrometric applications. It can help us to understand the physical conditions of the jet, such as the magnetic field  and the distance of the core 
from the base of the jet (\citealtads{Lobanov1998, Hirotani2005}).
Moreover, it affects the precise astrometric measurements performed by VLBI (\citealtads{Rioja2010, Porcas2009}).
The position of the core can depend on the activity state of the source
(e.g. \citealtads{Kudryavtseva2011}),
with recent studies revealing its variability
with time \citepads{Rushton2012, Paragi2013, Niinuma2015, Lisakov2017,Plavin2019}, an effect that, if not taken into account, can further affect the accurate astrometric observations of sources.

From a theoretical point of view, the relationship between core-shift and spectral characteristics was derived by \citetads{Konigl1981} for synchrotron emission by electrons with a power-law distribution of energy.
Using synchrotron emission equations and imposing optical depth equal to unity, the position, $l$, of the core at various frequencies results in a core-shift of $\nu^{-1/k_r}$, with $k_r$ 
depending on the spectral index, magnetic field, and particle density distribution \citepads{Konigl1981}.

In the present work, we aim to 
explore the core position in the jet parameter space, that is, as a function of 
frequency, magnetic field configuration, relativistic electron density, and jet inclination angle, $l \equiv l(\nu, B, \kappa, \eta)$, and to monitor the corresponding evolution of the related spectrum.
The importance of adding the inclination angle, $\eta$,
results from the shown anisotropy of jet emission with the inclination
\citepads{Zdziarski2016} and the evidence from
observations and simulations that the spectral index $\alpha$ can vary by changing the orientation of the jet 
(\citealtads{Fine2011, DiPompeo2012}, see their
Fig. 1).
We want to test here if the core position shows similar anisotropies to
those shown for  jet emission  and spectral index.
In particular, in this paper we analyse jets in microquasars.
Our code \citepads{Massi2014}, based on 
the synchrotron-emitting conical jet by \citetads{Kaiser2006}, provides us with the relevant information regarding the key parameters at different frequencies, that is, the position of the core ($l_{\tau=1}$) along the jet where $\tau=1$ and also the peak flux.
In Section~2, we describe the geometry of our synchrotron-emitting jet model. In Section~3, we present our results. Sections~3.1 and 3.2 describe the trend of the core position and spectral index, respectively, in response to variation in inclination angle, the electron density of the jet, and different magnetic field configurations. 
In Section~3.3, the relationship between core position and  spectral index 
 is analysed. 
Finally, we present our conclusions  in Section~4.




\section{Jet model}
Here, we describe the geometry that we adopt for the self-absorbed  stellar radio jet. The jet makes an angle $\eta$ with respect to the line of sight (LOS) and has a half-opening angle $\xi$ (see Fig.~\ref{fig:cone}). We choose $\eta$ such that $\xi < \eta \leq 75^{\degree}$ (Table~\ref{table:params}).
\citetads{Miller-Jones2006} calculated the half-opening angles of different X-ray binaries. Using the values from Table~1 provided by these latter authors, that is, for GRS~1915+105, Cygnus~X-3, GRO~J1655-40, XTE~J1550-564, H~1743-322 and 1RXS~J001442.2+580201, along with SS~433 \citepads{Marshall2013} and an updated value for Cygnus~X-1 \citepads{Tetarenko2019}, we find the average half-opening angle of eight different X-ray binaries to be $\xi \sim 3.5^{\degree}$. We use this value for our model.

\subsection{Geometry of the jet}
\label{sec:geometry}

\begin{figure}
\begin{center}
\includegraphics[width=0.7\columnwidth]{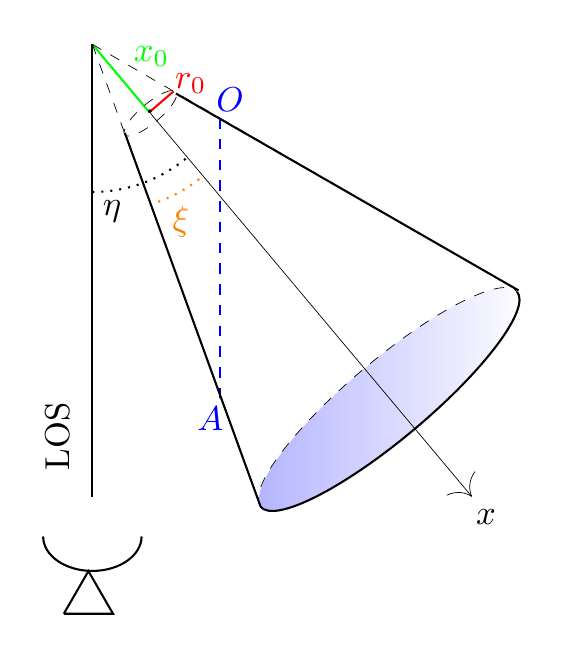}
\caption{Schematic representation of a jet with an inclination angle $\eta$ and a constant half-opening angle $\xi$ with $\textrm{tan } \xi = r_0/x_0$ (see text). The jet propagates along the x-axis with the jet base at $x_0$. Line segment $AO$ represents the path in a stratified jet as viewed by an observer along the LOS. The optical depths at $A$ and $O$ are $\tau=0$ and $\tau=\tau_{end}$, respectively.}
\label{fig:cone}
\end{center}
\end{figure}

\setcounter{table}{0}
\begin{table*}
\caption{Parameters used to study the characteristics of the adiabatic jet model.}
\centering
\begin{tabular}{l l l}
\hline
Symbol & Description &  Value \\ \hline
$\xi$ & Half-opening angle.  & $3.5^{\degree}$ \\
$\eta$ & Jet inclination angle.  & $5^{\degree}$--$75^{\degree}$ \\
$\kappa_0$ &  Parameter related to the relativistic electron density at the base of the jet  ($\rm{erg}^{0.8}$ $\rm{cm}^{-3}$).  &  6 and 0.6  \\
$x_0$ ($B_{||}$) &  Position of the base of the jet for parallel magnetic field (cm).  & 6.0 $\times 10^{12}$ \\
$x_0$ ($B_{\perp}$) &   Position of the base of the jet for perpendicular magnetic field (cm).  & 4.7 $\times 10^{12}$ \\
$r_0$ ($B_{||}$) &  Radius of the cone at the base  of the jet for parallel magnetic field (cm).  & 3.7 $\times 10^{11}$ \\
$r_0$ ($B_{\perp}$) &  Radius of the cone at the base of the jet for perpendicular magnetic field (cm). & 2.9 $\times 10^{11}$ \\
$B_0$ ($B_{||}$) & Magnetic field strength at the base of the jet for parallel magnetic field (Gauss). & 1.1  \\
$B_0$ ($B_{\perp}$) & Magnetic field strength at the base of the jet for perpendicular  magnetic field (Gauss). & 9.6  \\
$p$ & Power-law index.  & 1.8 \\
$D$ & Distance of the source (kpc). &  2.0  \\
\hline
\end{tabular}
\label{table:params}
\end{table*}

We follow the model for adiabatic jets provided by \citetads{Kaiser2006} and described in \citetads{Massi2014}. 
We consider a conical jet with the jet propagating along the x-axis.
All quantities are considered to be radially constant and change only along the x-axis. 
Any position along the x-axis is defined as $x=x_0l$ with the jet's base at position $x_0$, as shown in Fig.~\ref{fig:cone}.
The parameter $l$ is a dimensionless coordinate and is equal to unity at the base of the jet.
Adopting a distance of 2 kpc and the other values 
from Table~\ref{table:params}, any position along the x-axis when projected on the sky  plane is about $0.2~ l \sin \eta$ in mas. 
The radius of the cone is defined as $r=r_0l$, where $r_0$ is a constant scaling factor. The cone has a constant half-opening angle $\xi$ with $\textrm{tan } \xi = r_0/x_0$.
For our model, the base of the conical jet is located at $x_0 > 10^6 \textrm{ R}_{\rm{sch}}$ (see Table~\ref{table:params}), where $\textrm{ R}_{\rm{sch}}$ is the Schwarzschild radius. 
\citetads{Asada2012} discovered that the jet maintains a parabolic streamline over a range in size scale equal to $10^5 \textrm{ R}_{\rm{sch}}$, while further downstream the jet takes on a conical shape.
This result is consistent with the size of the acceleration/collimation region of a jet that is around $10^{4} \textrm{ R}_{\rm{sch}}$ following \citetads{Marscher2006conf}. 
Thus our jet model starts after the acceleration zone.

We parameterize the evolution of the magnetic field of the jet as $B=B_0 l^{-a_2}$, where $B_0$ is the magnetic field strength at the base of the jet and $a_2$ can take different values depending on the  
configuration of its magnetic field. From Table~1 of \citetads{Kaiser2006}, for a purely parallel magnetic field configuration, $a_2=2$ and for a 
purely perpendicular magnetic field configuration, $a_2=1$. 
We consider both of these cases in this work.

The number density of the relativistic electrons producing synchrotron radiation has a power-law distribution in energy,
\begin{equation}
N_{\rm{rel}} = \int^{E_{\rm{max}}}_{E_{\rm{min}}} \kappa E^{-p} \dif E,
  \label{eq:nrel}
\end{equation}
where $E$ is the energy of the electrons. The power-law index $p$ influences the flux density $S \propto \nu^{\alpha}$, with spectral index $\alpha$ related to $p$ as $\alpha = - (p-1)/2$ \citepads{Longair1994}. For our model, we use the parameters as given in Table~\ref{table:params}.
We assume the electron density distribution along the jet to evolve as $\kappa = \kappa_0 l^{-a_3}$, where $\kappa_0$ is a parameter related to the relativistic electron density at the base of the jet and $a_3 = 2(2+p)/3$ as derived by \citetads{Kaiser2006} from the mass conservation condition in the case of adiabatic losses. The spectral index range for optically thin emission is $ -1< \alpha < -0.2$ \citepads{Fender2001} which corresponds to $1.4<p<3$.
For $p=1.8$, $\kappa_0$ is expressed in units of [$\textrm{erg}^{0.8}$ $\textrm{cm}^{-3}$] throughout the paper.
The integral in the above equation is performed from $E_{\rm{min}}$ to $E_{\rm{max}}$, where $E_{\rm{min}}$ and $E_{\rm{max}}$ are the minimum and maximum energy of the electron distribution, respectively, and are related to the Lorentz factor, as discussed in Sect.~2.3.

\subsection{Synchrotron emission and optical depth}

In the framework of the jet model described above and developed by \citetads{Massi2014} for varying inclination angles, here we derive the flux density and opacity of the  jet model in terms of inclination angle $\eta$ and length of the jet, $l$. 
The perpendicular direction to the surface of the approaching jet makes an angle $90^{\degree}-(\eta-\xi$) with respect to the LOS, whereas it makes an angle $90^{\degree}-(\eta+\xi$) for the receding jet. Thus, the flux emanating from the approaching jet is given by 
\begin{equation}
  S_{\rm a}=\int^L_1  r_0 x_0 I_{\nu_{\rm a}}(\eta, l) {\sin (\eta- \xi) \over D^2} l {\textrm d}l,
\end{equation}
where $D$ is the distance of the source, and $I_{\nu_{\rm a}}$ is the intensity of the  approaching
jet. The integral is performed over the length of the jet from $l=1$ to $L$. We get the same expression for the receding jet flux ($S_r$), but the sign of the  angle is reversed. In this work, we consider a mildly relativistic jet with Doppler~factor~$\sim$~1 because we want to study the impact of opacity on jet properties due to the change in $\kappa_0$ and $\eta$. On the other hand, a greater Doppler factor increases the opacity (e.g. Eq.~10 in \citealtads{Finke2019}) and only enhances the studied effects, especially at small inclination angles.

For inclination angles $\xi<\eta < 90^{\degree}$, the observer views a stratified jet with varying plasma conditions along the LOS. For such a case, the total optical depth will have a contribution from each layer along the LOS. 
Therefore, using the radiative transfer equation, the intensity $I_{\nu}$ emanating from the surface of the jet and in the direction of the observer at an angle $\eta$ is given as  
\begin{equation}
I_{\nu }(\eta, l)= \int _{0}^{\tau_{\rm{end}}(l)} {J_{\nu} \over \chi_{\nu}}e^{- \tau' / \cos~\eta} ~{\dif} \large \left [{\tau' \over \cos~\eta} \large \right ].
\label{eq:3}
\end{equation}
From \citetads{Longair1994} and using Kaiser's notation, $J_{\nu}$ is the emissivity per unit volume, and $\chi_{\nu}$ is the absorption coefficient given by
\begin{equation}
\unit[ J_{\nu}= J_0 \nu^{-(p-1)/2} l^{-a_3-(p+1)a_2/2}]~~~~{{W {\textrm m}^{-3}{\textrm Hz^{-1}}}},
\end{equation}
and
\begin{equation}
\unit[ \chi_{\nu}= \chi_0 \nu^{-(p+4)/2} l^{-a_3-(p+2)a_2/2 }]~~~~ {{\textrm m^{-1}}},
\end{equation}
where,
\begin{equation}
  J_0=2.3~10^{-25}(1.3~ 10^{37})^{(p-1)/2}a(p)B_0^{(p+1)/2}\kappa_0,
\end{equation}
and
\begin{equation}
  \chi_0=3.4~10^{-9}(3.5~10^{18})^p b(p)B_0^{(p+2)/2}\kappa_0,
\end{equation}
with $a(p)$ and $b(p)$ being constants as given in \citetads{Longair1994}.
The optical depth is given by (as in \citealtads{Kaiser2006})
\begin{equation}
 \widetilde{\tau} (l)= \tau_0 l^C,
\label{eq:tau} 
\end{equation}
where $C = 1-a_3-(p+2)a_2/2$.
Using Table~1 from \citetads{Kaiser2006}, all possible values of the exponent 
in Eq.~\ref{eq:tau} will be negative, 
 that is, $C=-5.3$ and $C=-3.4$ for parallel and perpendicular magnetic field,
respectively.
Therefore, the optical depth is maximum at the base of the jet $l=1$ and is given by
\begin{equation}
  \tau_0= \widetilde{\tau}(l=1)={ \chi_0 x_0 \over -[1-a_3-(p+2)a_2/2]} \nu^{-(p+4)/2}.
\end{equation}

The integral in Eq.~\ref{eq:3} is performed over the optical depth along the LOS (along line segment $AO$ in Fig.~\ref{fig:cone}) from $\tau=0 \equiv \tau_A$ (if it is assumed that no emission or absorption occurs between the jet and the observer) to $\tau_{\rm{end}}(l)$, which represents the maximum optical depth of the jet, that is, until the position at~$O$. For an approaching jet, the optical depth $\tau_{\rm{end}}(l)$ is given as
\begin{equation}
\tau_{\rm{end}}(l) = \tau_0 l^C \large \left [\large \left (\frac{\tan \eta - \tan \xi}{\tan \eta + \tan \xi}\large \right)^C -1 \large \right].
\label{eq:10}
\end{equation}
See Eq.~A.5 in the Appendix of \citetads{Massi2014} for a comprehensive description.

To find the position along the jet where the optical depth attains a specific value $\tau =\tau'/\cos \eta$ (see Eq.~\ref{eq:3}), Eq.~\ref{eq:10} becomes $\tau_0 l^C f(\eta,\xi) = \tau \cos \eta$. Re-arranging the equation, we have
\begin{equation}
l(\tau) =  \left(\frac{\tau \cos \eta}{\tau_0 \textrm{ } \large \left [\large \left (\frac{\tan \eta - \tan \xi}{\tan \eta + \tan \xi}\large \right)^C -1 \large \right]}  \right)^{1/C}. 
\label{eq:14}
\end{equation}
As the core position is defined where the optical depth $\tau=1$, we can use Eq.~\ref{eq:14} to compute it along the jet ($l_{\tau=1}$) for any given frequency $\nu$, inclination angle $\eta$, relativistic electron density (dependent on the parameter $\kappa$; see Sect. 2.3), and  magnetic field configuration (dependent on the parameter $a_2$).

We check the consistency of our 
expression for $\tau$ with that of  Eq. 1
in  \citetads{Lobanov1998}, which was derived in the limit of a small jet 
opening angle.  
 This latter equation, with  
  $N=N_1 (r_1/r)^n$, $B=B_1 (r_1/r)^m$, $r_1=1$pc, and $\alpha=-0.5$, i.e.,  $s=1-2 \alpha=$2, 
  in the limit of  mildly relativistic flows, reads as 
\begin{equation}
\tau(r)_{Lobanov}= 6.1~10^{44}   N_1 B_1^2 r^{-(2m+n-1)}\nu^{-3} (\phi /sin (\theta)). 
\label{eq:14bis}
\end{equation}
From   Eq. 11,  our expression for $\tau$ is
\begin{equation}
\tau = {4.1~10^{26}}{ b(p) \kappa_0 x_0 \over - C \cos \eta}B_0^2 \nu^{-3} l^C \large \left [\large \left (\frac{\tan \eta - \tan \xi}{\tan \eta + \tan \xi}\large \right)^C -1 \large \right].
\label{eq:10four}
\end{equation}
This equation can be compared with Eq. 1 from \citetads{Lobanov1998} (here Eq. 12), for
 small $\xi$ and $\eta > \xi$ (\citealtads{Konigl1981}). In this limit, the angular part can be approximated as
\begin{equation}
 {[(1-2 {\xi\over \tan \eta})^C -1] \over \cos \eta}
\simeq {[(1-2 C {\xi\over \tan \eta} -1] \over \cos \eta}
=-2 C {\xi\over \sin \eta} . 
\end{equation}
Setting  the normalization distance $x_0=1$ pc and the exponent $p=2$, as in Lobanov,  with $b (p=2)= 0.269$ \citepads{Longair1994}, 
\begin{equation}
\tau = 6.8 ~10^{44} \kappa_0 B_0^2  l^{-(2a_2+a_3- 1)}  \nu^{-3} ( \xi / \sin \eta) .
\end{equation}
This expression is  the same 
as Eq. 12, taking into account  the correspondence between our  parameters and those of \citetads{Lobanov1998}, $a_2=m$, 
$a_3=n$, $\xi=\Phi$, and $\eta=\theta$ and the different approximations used
for synchrotron absorption;  in fact we used the formalization of \citetads{Longair1994}
instead of that of 
\citetads{Blumenthal1970} and  \citetads{Rybicki1979}, as in \citetads{Lobanov1998}.

\subsection{Model parameter $\kappa$ and relativistic electron density}
To understand the physical implications of the parameter $\kappa$, in this section, we calculate the relativistic electron density for Cygnus~X-1, as it is the best example of a system with a flat radio spectrum over the entire radio band until millimetre range \citepads{Fender2001}.
The range in energy for which these relativistic electrons exist is required to compute the relativistic electron density, and therefore, 
\begin{equation}
E_{\rm{min}} = \gamma_{\rm{min}} mc^2 \leq E \leq \gamma_{\rm{max}}mc^2 = E_{\rm{max}},
\end{equation}
where $E_{\rm{min}}$ and $E_{\rm{max}}$ is the minimum and maximum energy of the electron distribution, respectively and $\gamma$ is the Lorentz factor. As for a synchrotron process, $\gamma_{\rm{min}} >> 1$ \citepads{Dulk1985}, we set $\gamma_{\rm{min}}=10$.
Integrating Eq.~\ref{eq:nrel} gives 
\begin{equation}
N_{\rm{rel}} = \frac{\kappa}{1-p}(E_{\rm{max}}^{1-p}-E_{\rm{min}}^{1-p}).
\end{equation}
For $\gamma_{\rm{max}}>> \gamma_{\rm{min}}$, the number of relativistic electrons per cubic centimetre at the base of the jet ($l=1$) is given by
\begin{equation}
  N_{\rm{rel}}=\frac{\kappa_0}{p-1} (\gamma_{\rm{min}} mc^2)^{1-p}.
\end{equation}
For Cygnus X-1, the model of \citetads{Kaiser2006} with parameters from his Table~2 yield a relativistic electron density $N_{\rm{rel}} \approx 7\times 10^4 \rm{ cm}^{-3}$. 
We therefore chose $\kappa_0=6$ $\textrm{erg}^{0.8}$ $\textrm{cm}^{-3}$ for our jet model, which corresponds to the relativistic electron density $N_{\rm{rel}} \approx 7\times 10^4 \rm{ cm}^{-3}$ as for Cygnus X-1. 
We also study an intermediate case with $\kappa_0=0.6$ $\textrm{erg}^{0.8}$ $\textrm{cm}^{-3}$ with relativistic electron density $N_{\rm{rel}} \approx 7\times 10^3 \textrm{ cm}^{-3}$. In the latter part of Sect.~3, we generalise our study for $\kappa_0$= 0.06--60 $\textrm{erg}^{0.8}$ $\textrm{cm}^{-3}$.

\section{Results}
In this section, we use the parameters of Table 1 to analyse the characteristics of the above-described model of a self-absorbed jet as a function of the angle the jet axis makes with respect to the LOS and the injected relativistic electrons. We study the core-shift and spectral properties of the jet for both parallel and perpendicular magnetic field configurations. 

\subsection{Core-shift} 
\label{Section:core_freq}

\begin{figure}
\begin{center}
\includegraphics[width=0.47\columnwidth]{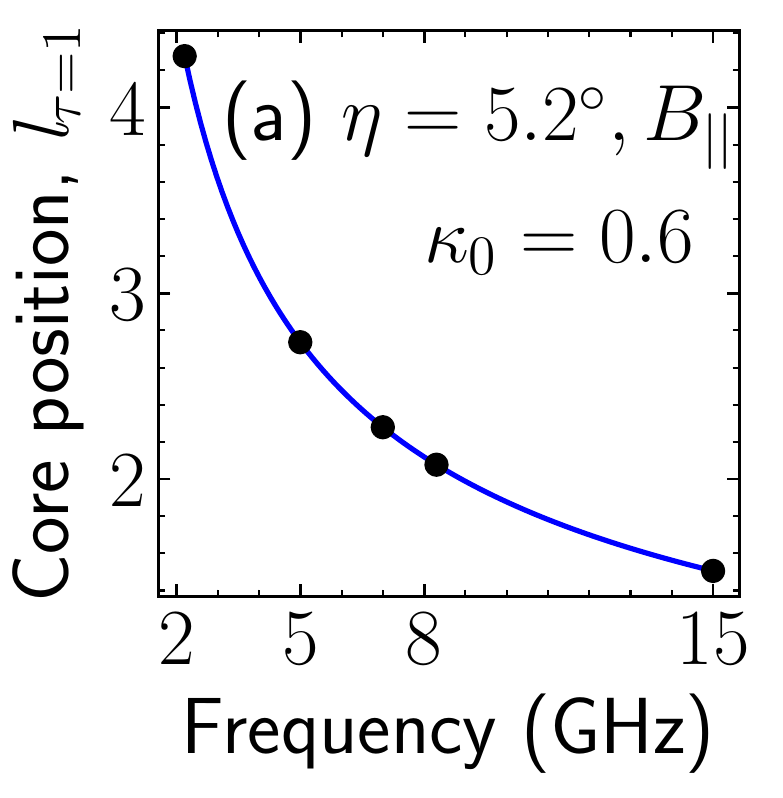} 
\includegraphics[width=0.47\columnwidth]{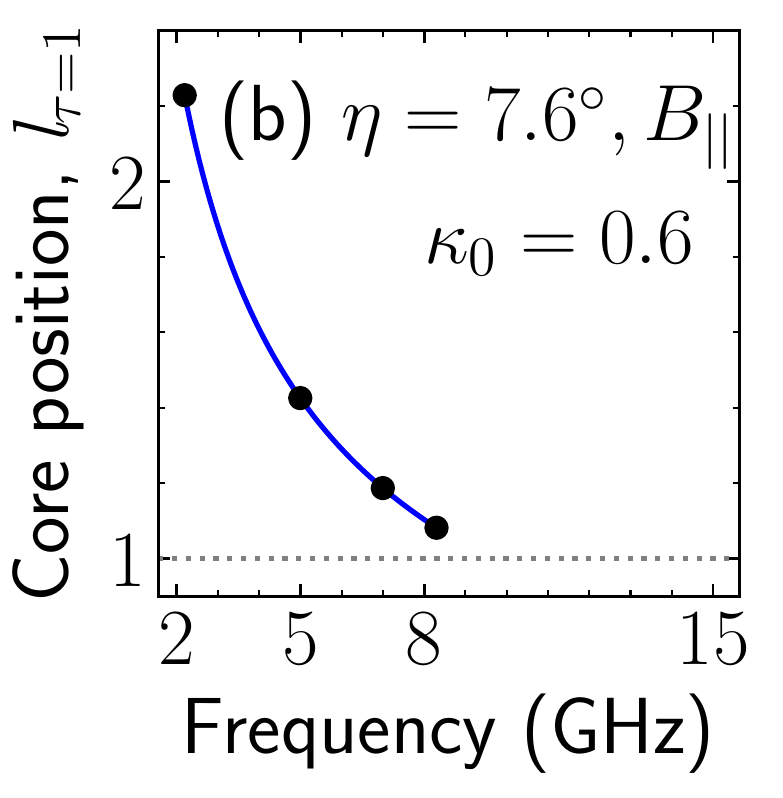}\\
\includegraphics[width=0.47\columnwidth]{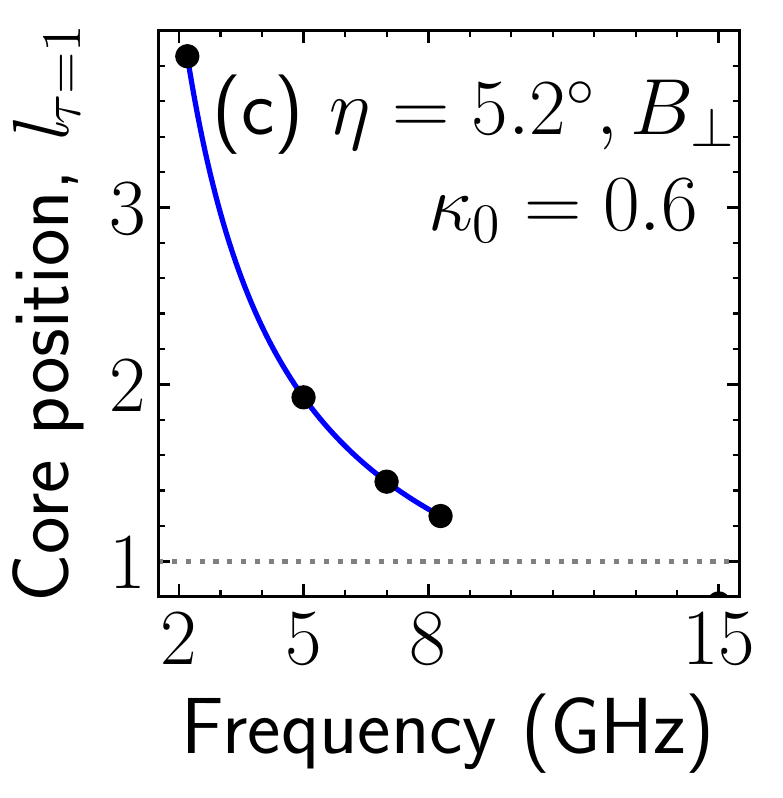}
\includegraphics[width=0.47\columnwidth]{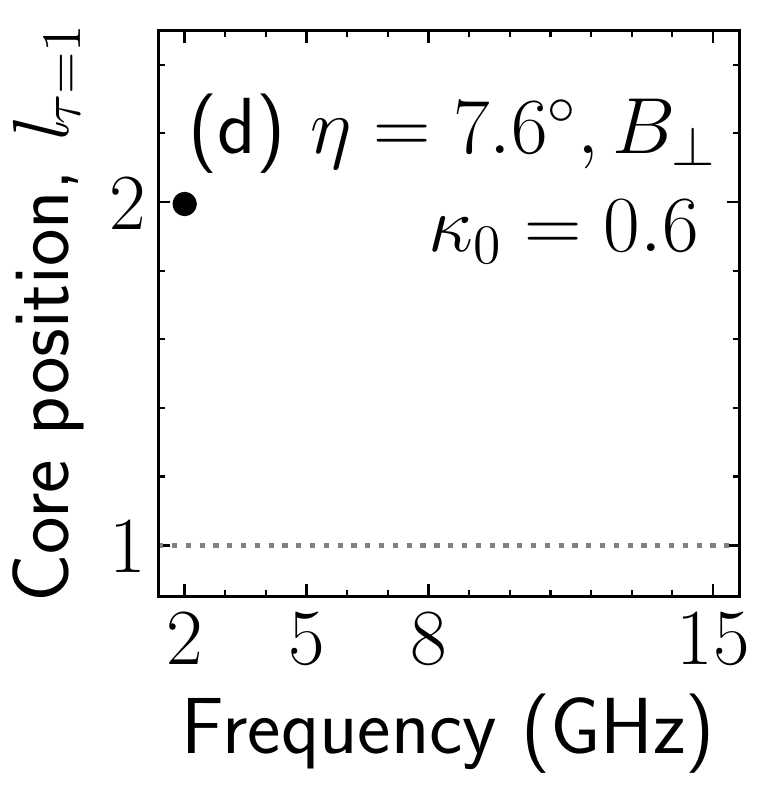}
\caption{Core positions (with reference to the base of the jet, $l=1$) 
for two different  inclination angles. 
Top panels: Parallel  magnetic field
configuration. Bottom panels: Perpendicular magnetic field configuration, with $\kappa_0=0.6$. The best-fit power-law function $l \propto \nu^{-1/k_r^{\prime}}$ is represented with blue lines.
 To determine the angular distance projected on the sky in milliarcseconds (mas)  the y-axis values must be multiplied by 
$0.2~\sin~\eta$ or $0.16~ \sin~\eta$ for parallel and  perpendicular B, respectively 
(for a distance of 2 kpc and the $x_0$  values of Table 1, see Sect. 2.1).}

\label{fig:core_shift}
\end{center}
\end{figure}

\begin{figure}
\begin{center}
\includegraphics[width=0.47\columnwidth]{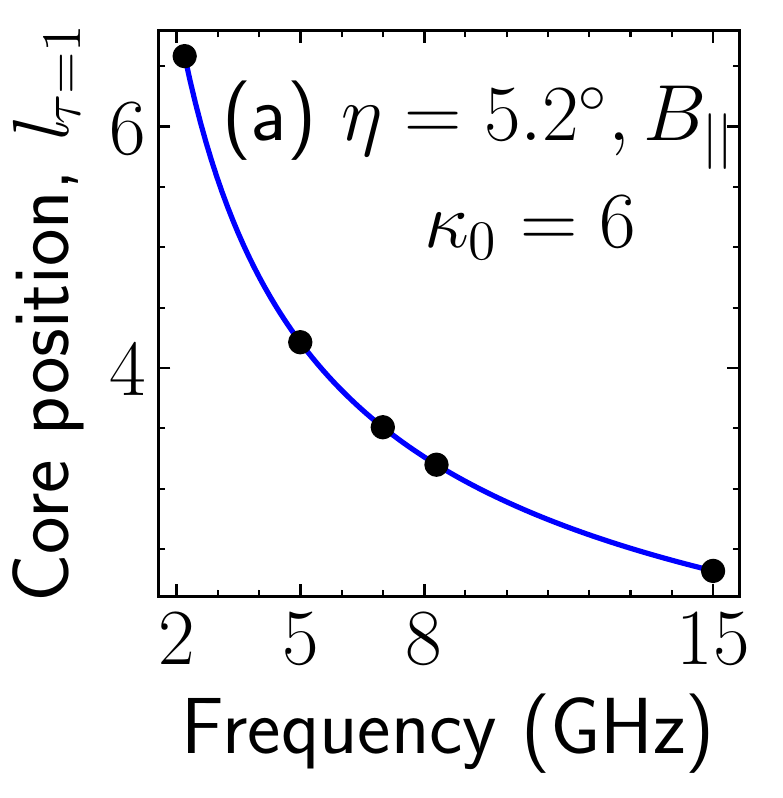} 
\includegraphics[width=0.47\columnwidth]{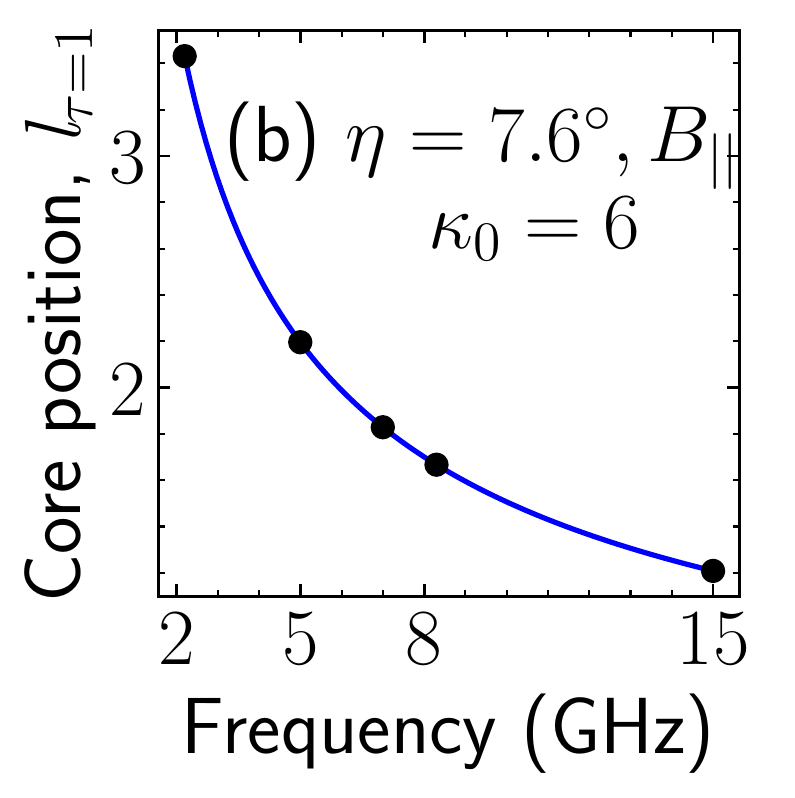}\\
\includegraphics[width=0.47\columnwidth]{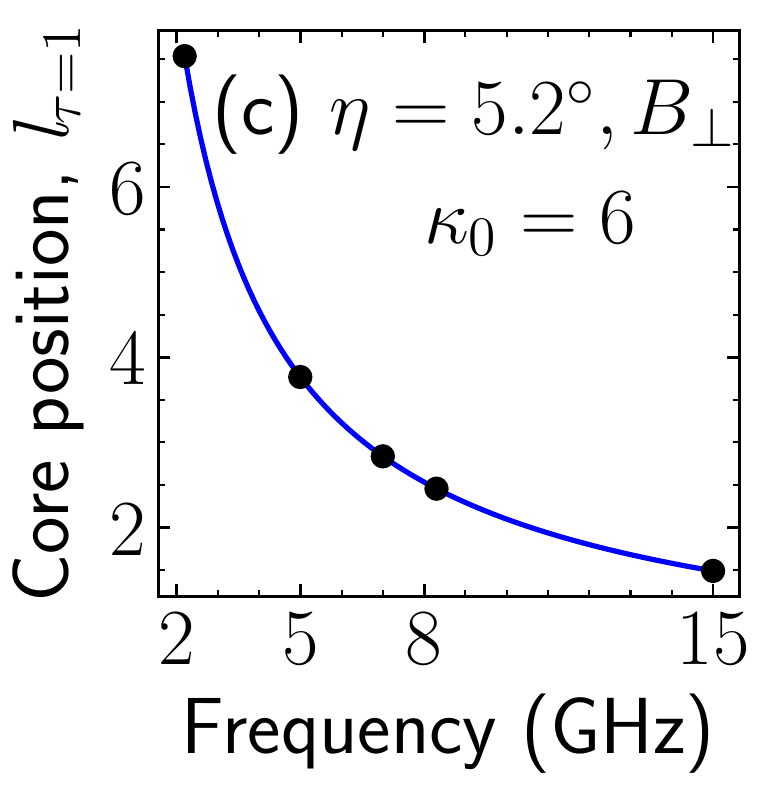}
\includegraphics[width=0.47\columnwidth]{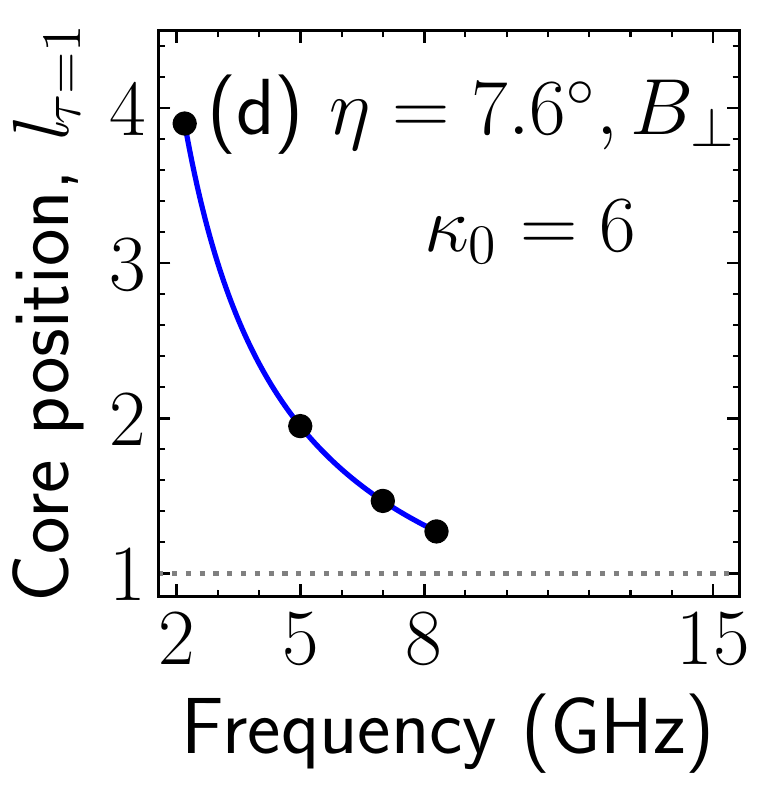}
\caption{Same as Fig.~\ref{fig:core_shift} but for $\kappa_0=6$.}
\label{fig:core_shift_6}
\end{center}
\end{figure}

Figures~\ref{fig:core_shift} and \ref{fig:core_shift_6} show the core position ($l_{\tau=1}$) for different
frequencies (2.2, 5, 7, 8.3, and 15 GHz) for parallel and perpendicular magnetic field configurations. The core positions are plotted for different inclination angles, $\eta = 5.2^{\degree}$ and $7.6^{\degree}$. 
Comparing Figs.~\ref{fig:core_shift}a with~\ref{fig:core_shift}c, we see that for a jet with the parallel magnetic field, the core is farther away from the base of the  jet  than that for a perpendicular magnetic field. For instance, the 5~GHz core position is at $l_{\tau=1} \approx 2.8$ for the parallel magnetic field (Fig.~\ref{fig:core_shift}a) and at $l_{\tau=1} \approx 2.0$ for the perpendicular magnetic field (Fig.~\ref{fig:core_shift}c).
Previous studies of a compact jet reveal that the core position is a function of the observing frequency, $l_{\tau=1} \propto \nu^{-1/k_r}$, where $k_r = ((3-2\alpha)a_2 +2a_3 -2)/(5-2\alpha)$. The power index $k_r$ depends on the magnetic field, particle density, and electron energy spectrum \citepads{Konigl1981}. 
Therefore, first, we fit a power-law function of the form $l_{\tau=1}= a \nu^{-1/k_r^{\prime}} $ (blue lines in Figs.~\ref{fig:core_shift},~\ref{fig:core_shift_6}) where $a$ is a constant depending on the fit. 
For the performed fits, for the parallel magnetic field, $k_r^{\prime} = 1.8$ and for the perpendicular magnetic field, $k_r^{\prime} = 1.2$, irrespective of the angle $\eta$. 
To calculate the theoretical values of $k_r$, we then use the spectral index values derived from the jet emission calculation for the given inclination angle.
For instance, for $\kappa_0=0.6$ and B$_{||}$, we have $\alpha=-0.13$ for $\eta=5.2^{\degree}$ (top panel of Fig.~\ref{fig:Spectral_index_qtau0pt6_par}). Using $a_2$ and $a_3$ from Sect.~\ref{sec:geometry}, we get $k_r=1.8$. Similarly, for $B_{\perp}$, we have $\alpha=-0.34$ for $\eta=5.2^{\degree}$ (top panel of Fig.~\ref{fig:Spectral_index_qtau0pt6_perp}) and we get $k_r=1.2$.
Thus, we see that the values of $k_r^{\prime}$ found by the fit of the derived core positions are equal to the theoretically calculated values of $k_r$  when using the derived spectral index for the same inclination angle.

\begin{figure}
\begin{center}
\includegraphics[width=1\columnwidth]{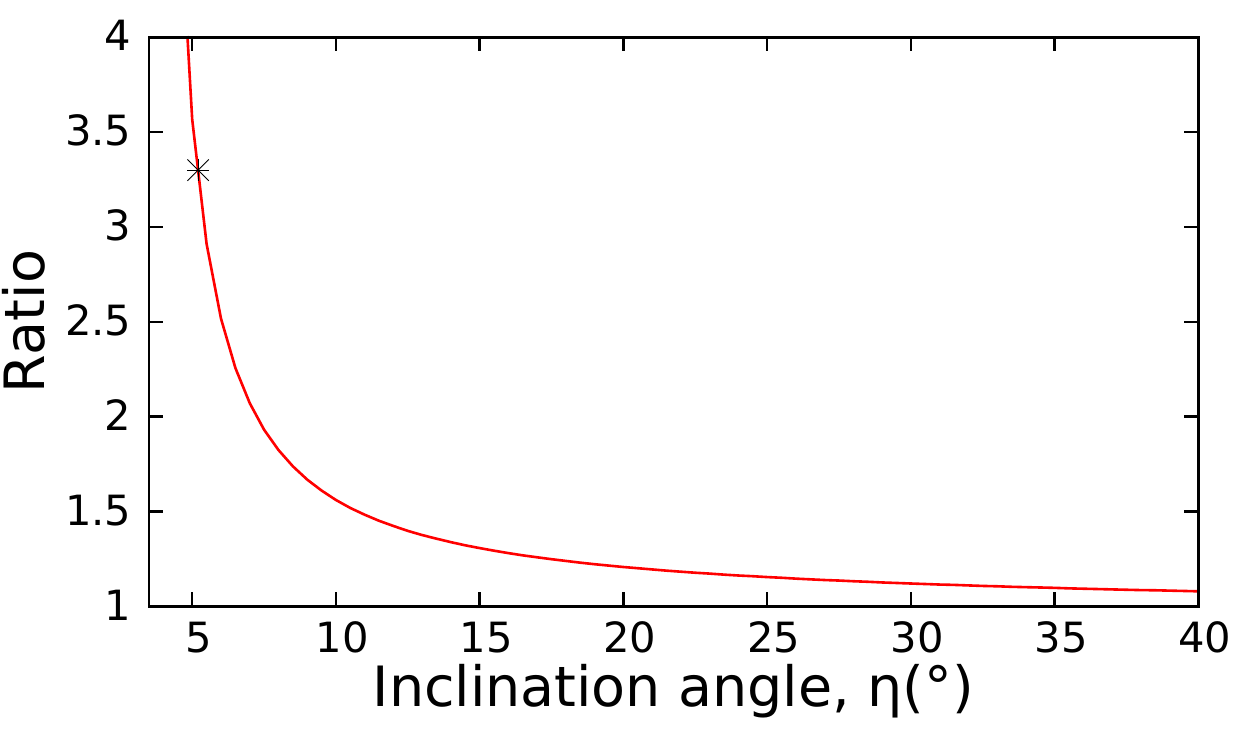}
\caption{Ratio of our  Eq. 11 with its approximation for
small $\xi$ and $\eta > \xi$, using  $\xi=3.5 ^{\degree}$ (see Sect. 2.2).
The symbol star indicates the ratio of 3.3 resulting for
$\eta=5.2^{\degree}$.}
\label{fig:scale}
\end{center}
\end{figure}

We also compare  our results with   Eq.~11 in \citetads{Lobanov1998}
where we set the bulk Lorentz factor equal to unity and  redshift $z=0$.
Assuming equipartition between jet particle
and magnetic field energies, the equation predicts the core shift $\Delta r$
at two frequencies  in a  jet  with synchrotron luminosity calculated
   using the model of \citetads{BlandfordKonig1979}.
To compare  our work  with Lobanov's equation  we chose
   the set of parameters 
of the perpendicular magnetic field (i.e. $a_2 =1, a_3 = 2.5, p = 1.8$) which are 
more compatible with Lobanov's set of parameters
(i.e. m=1, n=2, s=2).
For $\eta=5.2 ^{\degree}$,  $\xi=3.5^{\degree}$, using
  the values from Table 1 for perpendicular magnetic field,
namely $x_0=4.7 \times 10^{12}$cm and $B_0=9.6$ Gauss,
and a synchrotron
luminosity
  $L_{sync}=1.7 \times 10^{34}$ erg/sec
(following Eq. 23 in  \citealtads{BlandfordKonig1979}),
  Lobanov's Eq.~11 results in
a shift between 2 GHz and 8 GHz of   $\Delta r= 0.015$ mas.
In addition, one has to consider that 
our results compare with Lobanov's results
 for small $\xi$ and  $\eta > \xi$, as shown in Sect. 2.2.
Figure ~\ref{fig:scale} shows the ratio between our general equation (Eq. 11) on
which our plots of Figs. 2 and 3 are based,
and its approximation for  small $\xi$ and $\eta > \xi$.
Whereas the ratio tends to 1 for larger angles, at  $\eta=5.2^{\degree}$ the ratio is equal to 3.3.
Taking into account this correction factor of 3.3, the predicted $\Delta r= 0.015$
mas rises to 0.05 mas, well within the range of 0.04 mas $-$ 0.07 mas obtained from Figs. 2 c and 3 c.
Namely, the two values correspond to  $\kappa_0$ of 0.6 (Fig. 2 c)
and 6.0 (Fig. 3 c),
and from the equipartition the value for  $\kappa_0$ from $B_0 = 9.6$ Gauss is  $\kappa_0
= 3.6$.

\begin{figure}
\begin{center}
\includegraphics[width=1\columnwidth]{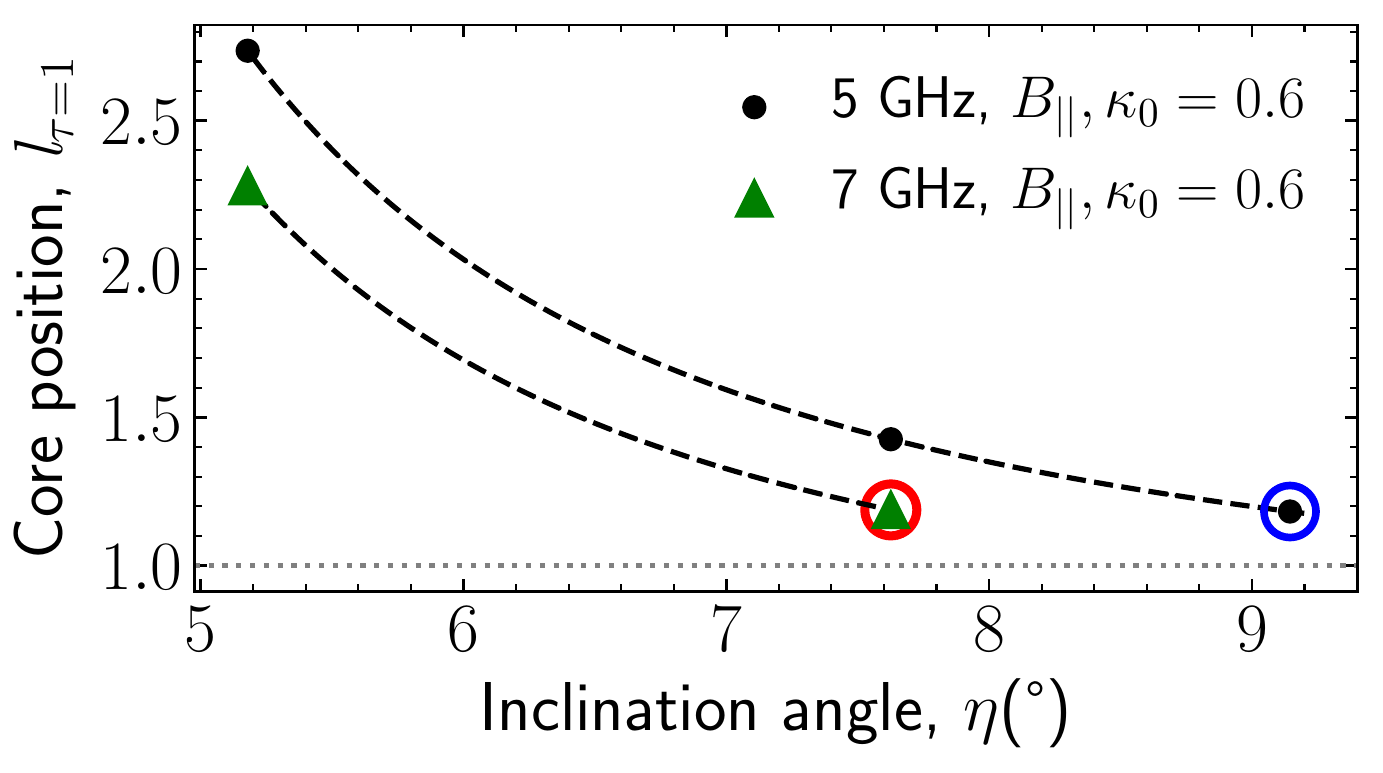}
\caption{Core position as a function of the inclination angle for 5 GHz and 7~GHz with 
parallel magnetic field configuration and $\kappa_0=0.6$. 
The dashed lines represent an exponential decay in the positions as the angle increases, given by $l \propto e^{q/\eta}$. The dotted line represents the jet base.
The optical depth becomes $\tau<1$ all along the  conical part of the jet  
for angles greater than 9.1$^{\degree}$ (blue circle) and 7.6$^{\degree}$ (red circle) for 5 GHz and 7 GHz, respectively. This last inclination angle where the core is present in the conical jet is plotted for all frequencies and different $\kappa_0$ in Fig.~\ref{fig:kappaVSeta_Core}.}
\label{fig:CoreVsEta_5ghz}
\end{center}
\end{figure} 
\begin{figure}
\begin{center}
\includegraphics[width=1\columnwidth]{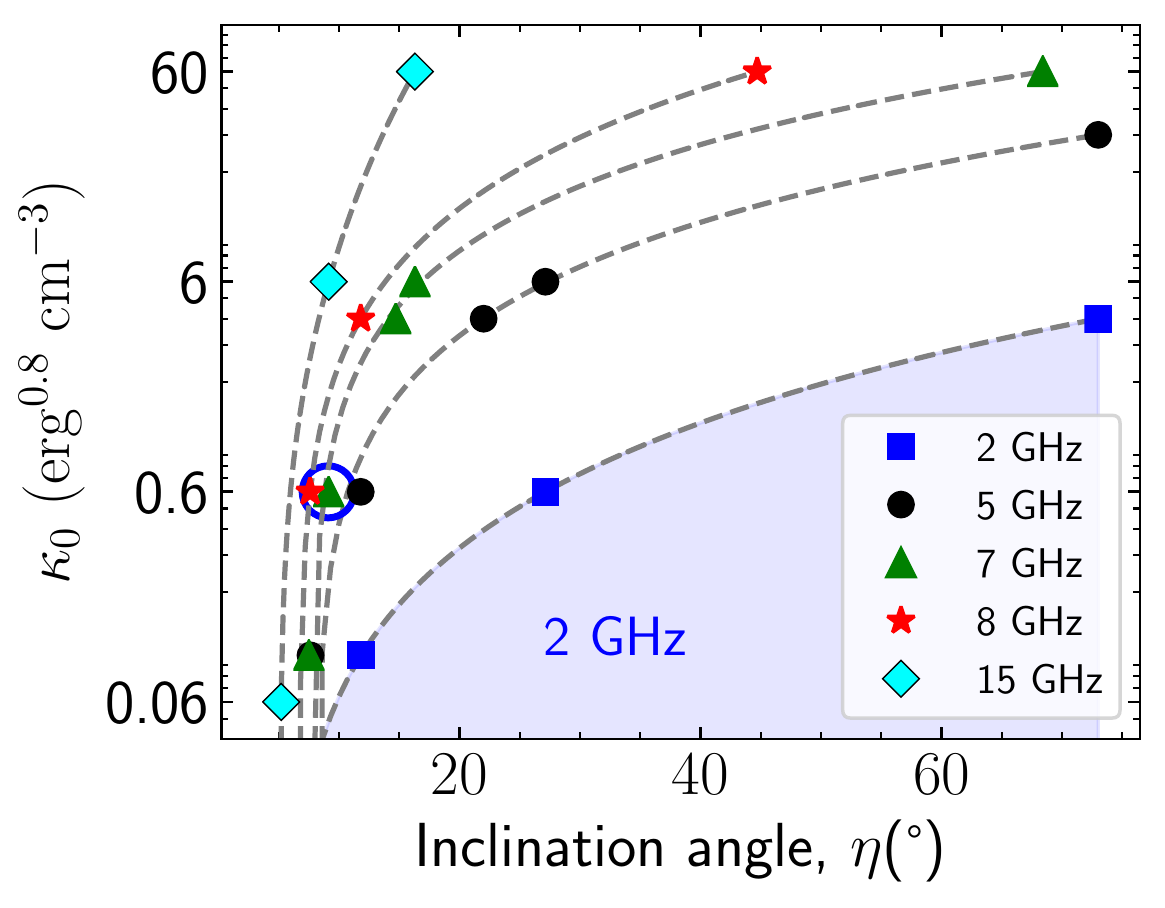}\\
\includegraphics[width=1\columnwidth]{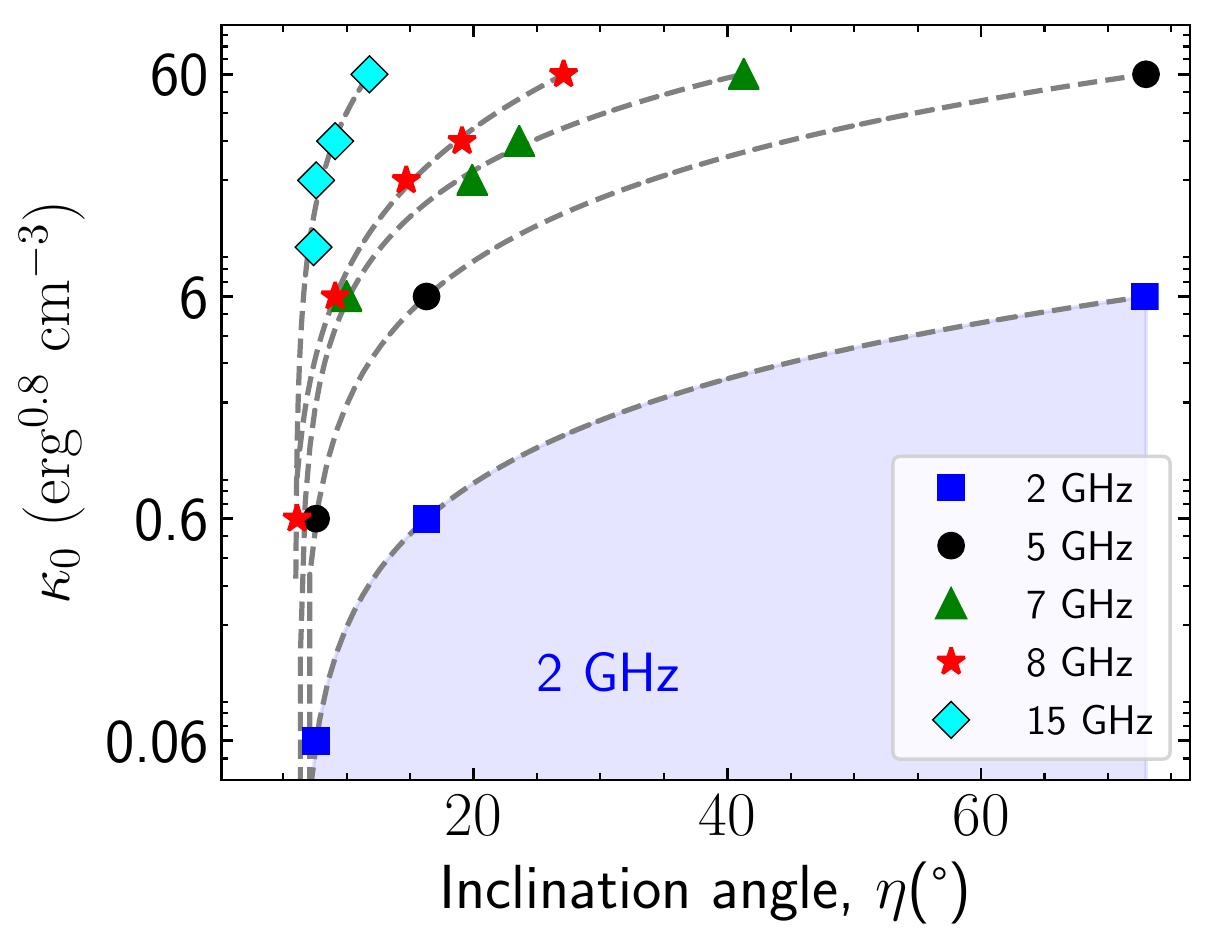}
\caption{Last inclination angle in the ($ \eta, \kappa_0$)--plane where the core is still present in the conical jet for the given frequencies (2, 5, 7, 8, and 15 GHz), before 
 moving  below the base.
The curve divides the ($ \eta, \kappa_0$)--plane into two distinct regions: 
presence of core along the 
conical jet, i.e. $l_{\tau=1} \geqslant 1$ (points on and area above the curve) and 
core  moved below the base of the jet 
$l_{\tau=1} < 1$ (area under the curve) for a given frequency.
As an example, here we trace the area under the curve for 2 GHz (blue shaded region). Y-axis is in log scale. Top and bottom panels show results for parallel and perpendicular magnetic field configurations, respectively.}
\label{fig:kappaVSeta_Core}
\end{center}
\end{figure}

In addition to verifying the known core-shift relationship with frequency and agreement of our findings with theoretical predictions, we obtain a new result.
Figure~\ref{fig:core_shift} shows that for the same magnetic field configuration, increasing the inclination angle $\eta$ shifts the core position of a given frequency closer to the base of the jet. For example, for the parallel magnetic field in Figs.~\ref{fig:core_shift}-a,b, the core position for 5 GHz emission is at $l_{\tau=1}\approx 2.8$ for $\eta=5.2^{\degree}$ and decreases to $l_{\tau=1}\approx 1.4$ for $\eta=7.6^{\degree}$. 
Similarly, for perpendicular magnetic field (see Figs.~\ref{fig:core_shift}-c,d), let us consider again the core position for 5 GHz emission, which is at $l_{\tau=1}\approx 2.0$ 
for $\eta=5.2^{\degree}$ but  for $\eta \geq 7.6^{\degree}$ 
 the optical depth becomes $\tau<1$ all along the  conical part of the jet. 
This shows that the core-shift 
depends not only on the frequency but also on the jet's inclination angle.

Another important result from our analysis is shown in Fig.~\ref{fig:core_shift_6}, which shows the changes in core position for a higher relativistic electron density, that is, for $\kappa_0=6$. We see that for higher relativistic electrons, the core moves downstream, that is, away from the base of the jet. Let us revisit the core position for 5~GHz. In Fig.~\ref{fig:core_shift}a, the core position for 5 GHz is at $l_{\tau=1}\approx 2.8$ for $\kappa_0=0.6$, whereas it increases to $l_{\tau=1}\approx 4.2$ for $\kappa_0=6$ as shown in Fig.~\ref{fig:core_shift_6}a.

The change in the core position for different inclination angles $\eta$ is shown in Fig.~\ref{fig:CoreVsEta_5ghz} for 5 GHz and 7 GHz for the parallel magnetic field. The position of the core can be seen to decay with increasing inclination angle. The last angle for 5 GHz where the core is present in the
conical jet is $\eta \approx 9.1^{\degree}$ and is shown by a blue circle. For larger angles,   the emission of 5 GHz becomes optically thin
all along the conical jet.
This holds true for the particular case of $\kappa_0$=0.6. As shown in Fig.~\ref{fig:kappaVSeta_Core}, at larger electron densities,
the core  persists  at much larger angles.
In Fig.~\ref{fig:CoreVsEta_5ghz},  
 for 7 GHz, the core is present in the conical jet until $\eta \approx 7.6^{\degree}$. 
We performed a fit with an exponential function of the form,
$l \propto e^{q/\eta}$, as shown by dashed lines where $q$ is a constant depending on the fit.

In Fig.~\ref{fig:kappaVSeta_Core}, we study the presence of the core in the 
conical jet for each frequency and different relativistic electron densities (ranging from $\kappa_0=$ 0.06--60.0).
For a given frequency, each position marked in the ($\eta, \kappa_0$)--plane refers to the last inclination angle for which the core is present for that frequency, before  moving below the base of the jet. This way, the blue circle at ($\eta, \kappa_0$)~=~($9.1^{\degree}, 0.6$) for 5 GHz represents the same condition as the blue circle in Fig.~\ref{fig:CoreVsEta_5ghz}. The dependence of $\kappa_0$ on $\eta$ is fitted with a power-law function for all frequencies. 
The curves divide the ($\eta, \kappa_0$)--plane into two distinct regions:  
any position on and above the curve represents the presence of the  core in the conical jet
for a given frequency; 
for example, for 2 GHz, it is the region above the blue shaded area. 
For instance, as shown in the top panel of Fig.~\ref{fig:kappaVSeta_Core}, for the parallel magnetic field configuration and $\kappa_0=0.06$, for a frequency of 2~GHz, the core is present until an inclination angle of $\eta \approx 9.1^{\degree}$. 
On the other hand, for 15 GHz, the core is present until $5.2^{\degree}$ at this relativistic electron density. To observe the core at a greater inclination angle, we would need to increase the number of relativistic electrons in the jet. 
Therefore, for the core to be present  in the conical jet
for 2 GHz at $\approx 27^{\degree}$, we need an increased electron density with $\kappa_0= 0.6$.
The same applies 
to  the perpendicular magnetic field configuration (bottom panel of Fig.~\ref{fig:kappaVSeta_Core}).

Very long baseline array (VLBA) astrometry of the microquasar LS~I~+61$^\circ$303  \citepads{Wu2018}
can be better understood in light of our results.
A  sequence of ten observations  was spaced 3-4 days apart 
and  the  core of the precessing jet was expected to follow a
regular path  as well, that is, with regularly spaced astrometric points.
Instead, VLBA astrometry \citepads{Wu2018}
 shows that the positions of the cores are displaced from the jet 
base, with the largest offsets for those observations  corresponding 
to the smallest inclination angles.
A future quantitative comparison of  our determined angular distances  
projected on the sky with observed astrometry will make it possible 
to constrain important parameters such as $\kappa_0$ and  $B_0$ for LS~I~+61$^\circ$303 
and other microquasars.

\begin{figure*}
\begin{center}
\includegraphics[width=2\columnwidth]{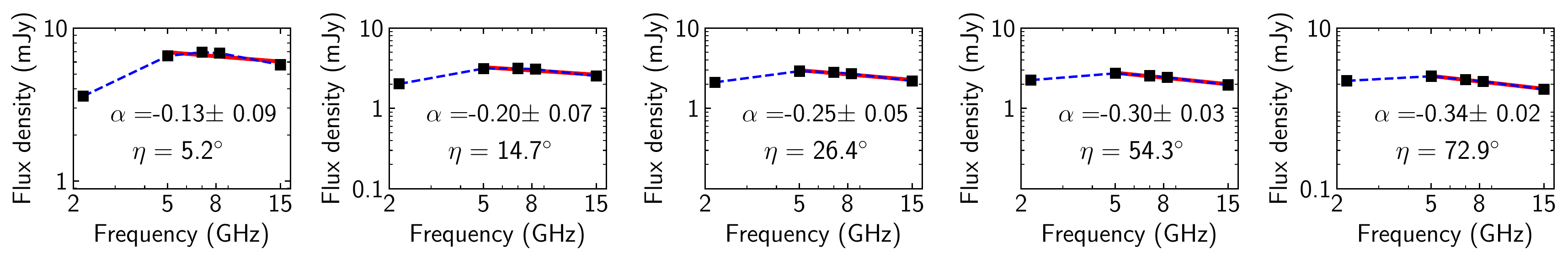}\\
\includegraphics[width=2\columnwidth]{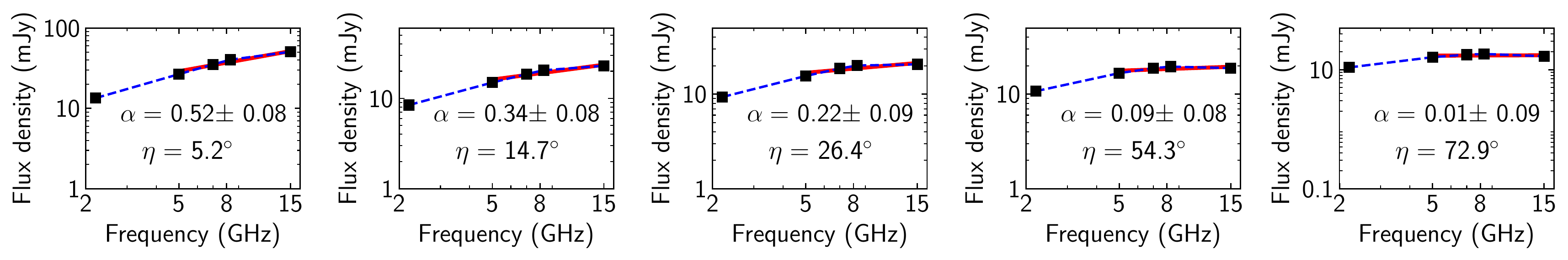}
\caption{Flux as a function of frequency for different inclination angles for a parallel magnetic field configuration. Both axes are in log scale. The blue dashed lines connect the model data and the red lines show the spectral index fit. Top panel: With $\kappa_0 = 0.6.$ Bottom panel: With $\kappa_0 = 6$.}
\label{fig:Spectral_index_qtau0pt6_par}
\end{center}
\end{figure*}

\begin{figure*}
\begin{center}
\includegraphics[width=2\columnwidth]{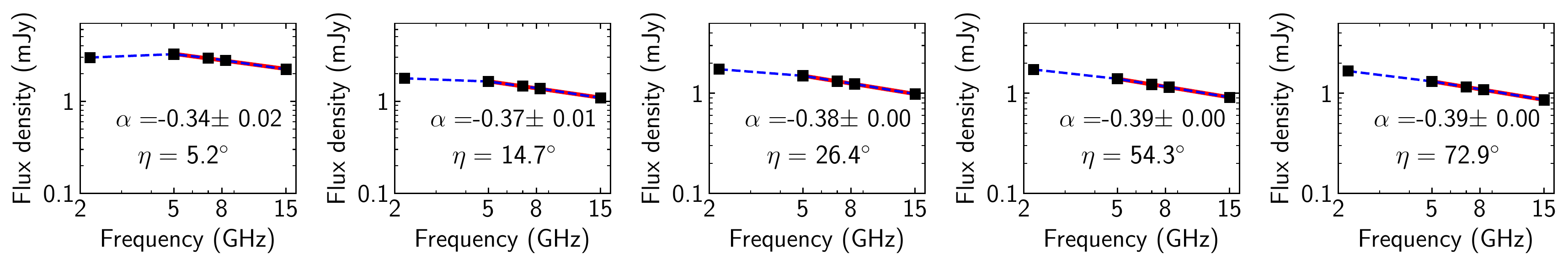}\\
\includegraphics[width=2\columnwidth]{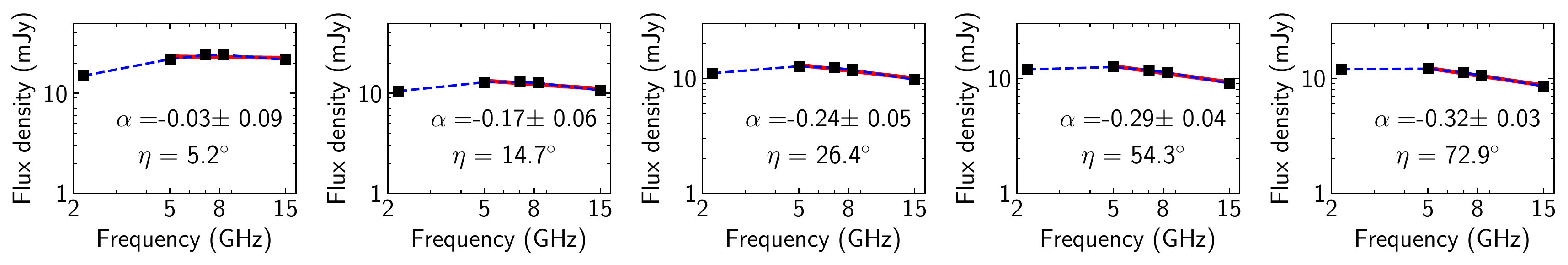}
\caption{Same as Fig.~\ref{fig:Spectral_index_qtau0pt6_par}, but for a perpendicular magnetic field configuration.}
\label{fig:Spectral_index_qtau0pt6_perp}
\end{center}
\end{figure*}

~\\
~\\
\subsection{Spectral index analysis}
\label{section:Spectral_index}
Figures \ref{fig:Spectral_index_qtau0pt6_par} and~\ref{fig:Spectral_index_qtau0pt6_perp} show the spectra of the jet for different inclination angles for parallel and perpendicular magnetic field cases, respectively. 
We show the power-law fit required to get the spectral index $\alpha$ only for higher frequencies, i.e. excluding 2~GHz. This is because the 2~GHz core survives the largest inclination angle, and including it would bias the fit, resulting in large uncertainties for the spectral index. For instance, as shown by the blue dashed line, the spectra between 2~GHz and 5~GHz are inverted, contrary to the spectra between other frequencies (top panel of Fig.~\ref{fig:Spectral_index_qtau0pt6_par}). 

Considering the spectra from 5--15~GHz in the top panel of Fig.~\ref{fig:Spectral_index_qtau0pt6_par} for $\kappa_0=0.6$, the spectral index indicates optically thin emission ($\alpha<0$) at all inclination angles while becoming more negative for increasing angles. 
For the same electron density, but for the perpendicular magnetic field (top panel of Fig.~\ref{fig:Spectral_index_qtau0pt6_perp}), the emission is again optically thin. However, the spectral index is more negative than for the corresponding angles of the parallel magnetic field. 
When we increase the relativistic electron density to $\kappa_0 = 6$ (bottom panel of Fig.~\ref{fig:Spectral_index_qtau0pt6_par}), the spectra become inverted ($\alpha>0$) and only become flat around $73^{\degree}$. For the perpendicular case (bottom panel of Fig.~\ref{fig:Spectral_index_qtau0pt6_perp}), the spectra are initially flat 
at $5.2^{\degree}$ and later become optically thin. Our analysis shows that if we increase the relativistic electron density, the spectra become flat or even  inverted.
Thus we see that the spectral index changes significantly depending on the inclination angle, magnetic field configuration, and relativistic electron density, similar to core position changes as shown in the previous section.

We  compare our results with 6 yr of radio  observations of
 LS~I~+61$^\circ$303.
Spectral index data for LS~I~+61$^\circ$303 are shown in Fig. 5 of
\citetads{MassiKaufman2009}. The spectral index versus orbital phase presents a recurrent pattern:
$\alpha \geq 0$ twice along the orbit, during the maximum of the long-term
flux density modulation 
where $\Theta=0.0-0.1$, and $\Theta=0.9-1.0$, i.e. for ejections at small inclination angles \citepads{Massi2014}.
The trend changes and evolves  to $\alpha < 0$ along the orbit during the minimum 
of the long-term flux density modulation where $\Theta=0.4-0.5$, i.e. for ejections at the largest inclination
angles \citepads{Massi2014}.

\begin{figure}
\begin{center}
\includegraphics[width=1\columnwidth]{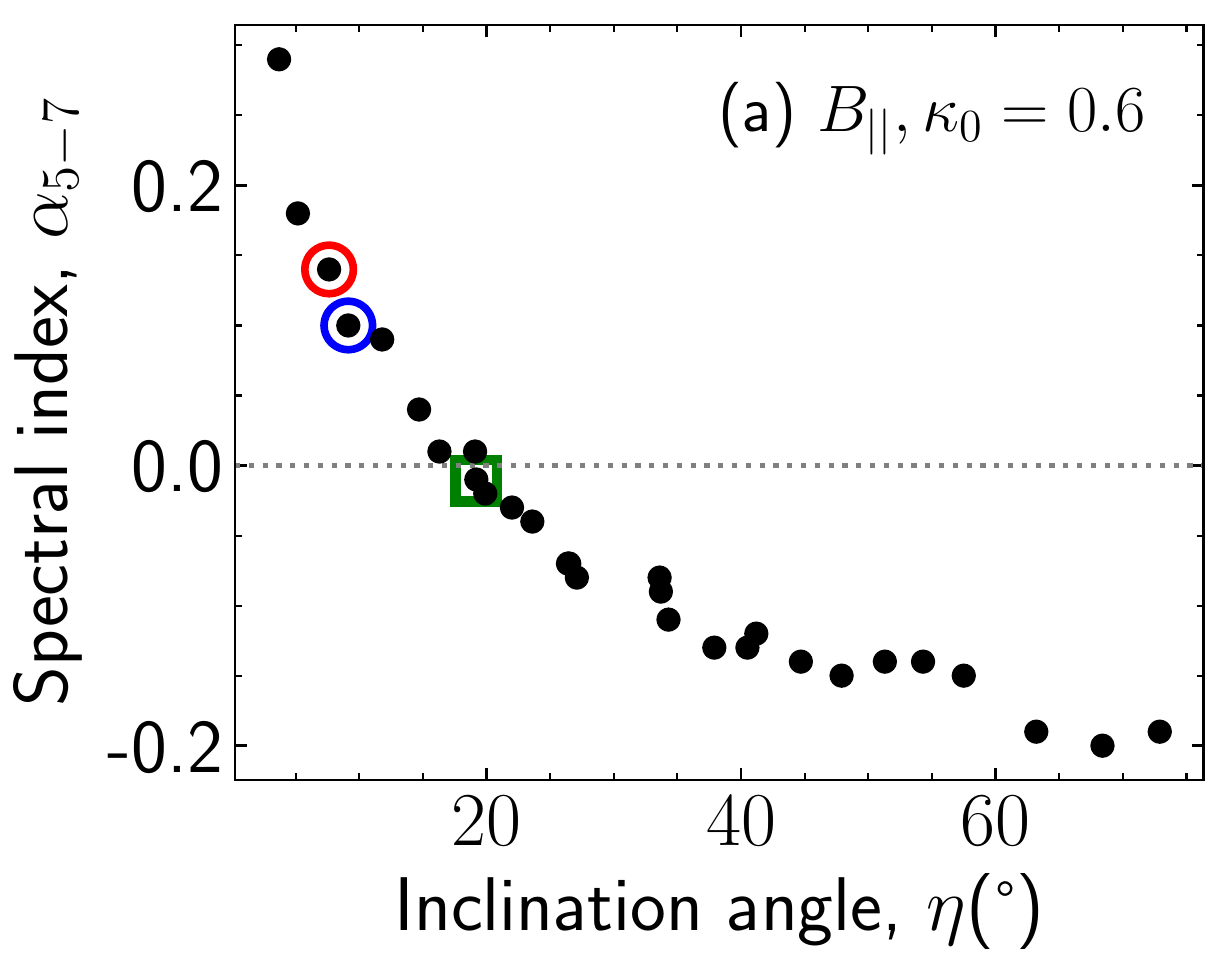} 
\includegraphics[width=1\columnwidth]{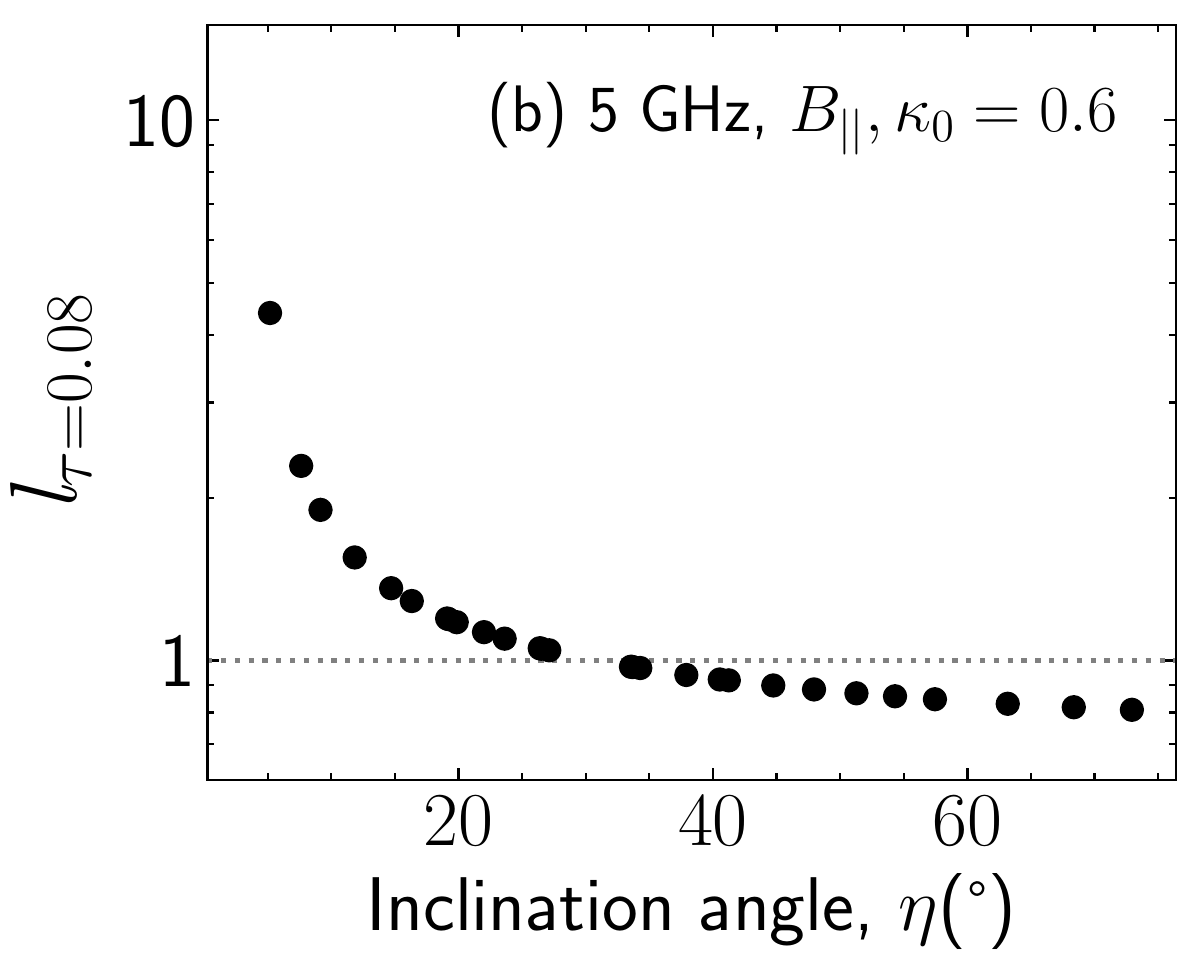}
\includegraphics[width=1\columnwidth]{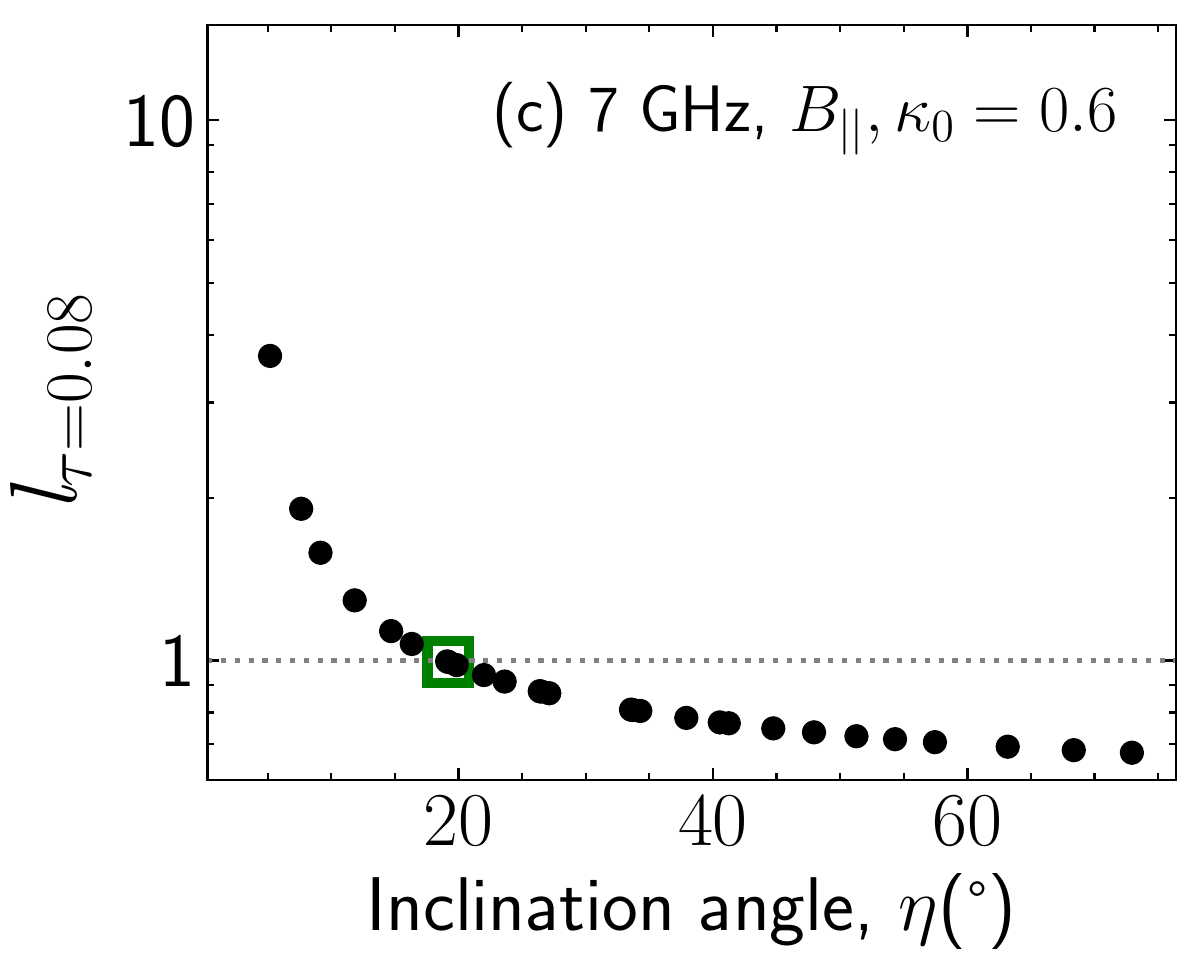}
 \caption{(a) Spectral index between 5 GHz and 7 GHz as a function of the inclination angle  for $\kappa_0=0.6$. The green square represents the angle where the emission becomes optically thin. Red and blue circles are the same as in Figs.~\ref{fig:CoreVsEta_5ghz} and~\ref{fig:kappaVSeta_Core}. (b),(c) Positions of the jet surface where optical depth  $\tau=0.08$ for 5 GHz and 7 GHz, respectively,
with $\kappa_0=0.6$. $l_{\tau=0.08}=1$ represents the position on the jet after which the emission becomes optically thin all along the jet, i.e.  when $\tau<<1$.}
\label{fig:alphaVsEta}
\end{center}
\end{figure}
~\\
\subsection{Relationship between core position and spectral index}

We analysed the core positions and spectral indices in previous sections by changing the inclination angles and relativistic electron density. In this section, we study their mutual relationship.

In Fig.~\ref{fig:alphaVsEta} a, we show the change in the spectral index between 5 GHz and 7 GHz for different inclination angles. For $\kappa_0=0.6$, we see that the spectrum is inverted for angles below $\eta=19^{\degree}$ and becomes optically thin from $\eta=19.1^{\degree}$ (marked by the green square).
As noted above, the core position at the jet base, $l_{\tau=1}=1$, is present until $\eta=9.1^{\degree}$ for 5 GHz (blue circle in Figs.~\ref{fig:CoreVsEta_5ghz} and~\ref{fig:alphaVsEta}a) and until $\eta=7.6^{\degree}$ for 7 GHz (red circle in Fig.~\ref{fig:CoreVsEta_5ghz}). This is different from the expected angle of $\eta=19.1^{\degree}$. 
This means that even though the core is not present, $\alpha$ still remains positive near the base of the jet.
For the spectrum to evolve from inverted (i.e. $\alpha \geqslant 0$) to optically thin (i.e. $\alpha<0$), the optical depth must fulfil the condition $\tau << 1$ all
along the jet, including at the base of the jet.
We quantified this value $\tau << 1$ with our code.
For that purpose, we performed several iterations of the code assuming different values for $\tau = \tau'/\textrm{cos } \eta $ (see Eq.~3). At each iteration, we checked for the spectral index transition from $\alpha \geqslant 0$ to $\alpha < 0$. We find that for our jet model, the transition occurs when $\tau'/\textrm{cos } \eta=0.08$ at the base of the jet ($l=1$), confirming that the condition for an optically thin jet is $\tau = \tau'/\textrm{cos } \eta << 1$.
This implies that the transition of $\alpha$ to negative values on the jet surface does not happen at $l_{\tau=1}=1$, that is, at the core at the base of the jet,
but at $l_{\tau=0.08} = l(\tau'/\textrm{cos } \eta=0.08)=1$, that is, when the jet becomes completely optically thin until the base. Applying the condition $\tau=0.08$ in Eq.~\ref{eq:14}, for $\kappa_0=0.6$, the optically thin emission begins at $\eta=19.1^{\degree}$ (marked by the green square) as shown in Figs.~\ref{fig:alphaVsEta}-b,c.

\section{Conclusions}

We studied the synchrotron-emitting conical, self-absorbed jet initially proposed by \citetads{BlandfordKonig1979}, later developed by \citetads{Kaiser2006}, and finally
 modified  to vary the inclination angle by \citetads{Massi2014}.
The aim of this study is to explore the core position, $l$, as a function of 
frequency, magnetic field alignment, relativistic electron density, and jet inclination angle.
 
First, we checked  that our derived  expression for the optical depth,
that is, the fundamental equation in our analysis,
 is  equal to  the commonly used Eq. 1
in  \citetads{Lobanov1998}.
From our jet model 
results, we find  that without changing the inclination angle of the jet or without changing the density of the relativistic electrons, the position of the core for a higher frequency is closer to the base of the jet than for lower frequencies, as expected in light of the synchrotron self-absorption effects \citepads{Konigl1981}. 
We then checked the consistency of  our results, specifically paying attention to    
the obtained position of the cores at various frequencies
with the derived  core-frequency relationship of the jets, $l \propto \nu^{-1/k_r}$ 
by \citetads{Konigl1981}
and also with the derived shift of the core position between two frequencies
calculated  by Eq. 11 in \citetads{Lobanov1998}.

Finally, we proceeded to  study the core position in the jet at different inclination angles 
and electron densities.
Our results in Figs.~\ref{fig:core_shift} and~\ref{fig:core_shift_6} show that for a given frequency, 
the greater the inclination angle,  the closer the core is to the base of the jet. 
This implies  that the core position can vary in an irregular way 
if the jet is precessing.   
Therefore, the  deviations from a  regular path are  expected, as indeed traced by astrometric observations  
by \citetads{Wu2018}
which are discussed here in Section 3.1. 
As shown in Fig.~\ref{fig:core_shift_6}, the core is more displaced from the jet base for higher electron densities.
This is consistent with the observations where increased relativistic electron density due to flares shows a displacement of the core position downstream from the jet \citepads{Niinuma2015, Lisakov2017, Plavin2019}. 
Figure~\ref{fig:kappaVSeta_Core} summarise the results of variations in inclination angle, electron density, and magnetic field configuration. 
The curves in the ($\eta, \kappa_0$)--plane show where 
for a given frequency 
the  conical 
jet becomes entirely  optically thin. 
The corresponding spectra, functions of the same parameters, are shown in Figs.~\ref{fig:Spectral_index_qtau0pt6_par} and~\ref{fig:Spectral_index_qtau0pt6_perp}.
Comparing the spectral index for different relativistic electron densities, we see that an increase in electron density causes the spectra to become less steep or even  inverted. Such results have been confirmed by observations of radio flares (e.g. \citealtads{Lisakov2017}).
Moreover, 
we see that
the spectral index varies by changing the jet orientation,
with a steep spectrum at large inclination angles getting flatter  as the angle decreases
(top panel of 
Fig.~\ref{fig:Spectral_index_qtau0pt6_par} 
and bottom panel of 
Fig.~\ref{fig:Spectral_index_qtau0pt6_perp}),
or a flat spectrum at large inclination angles
becoming inverted as the angle decreases
(bottom panel of 
Fig.~\ref{fig:Spectral_index_qtau0pt6_par}).
These results for the relationship between spectral index and inclination angle are consistent with 
observations, 
as discussed in Section 3.2.
Moreover,  simulations (Fig.~1 in \citealtads{DiPompeo2012})
 show the same trend between $\alpha$ and $\eta$ as in Fig.~\ref{fig:alphaVsEta}.  
In Fig.~\ref{fig:alphaVsEta}a, the spectral index at low inclination begins with a high positive value (inverted spectrum). 
Increasing the value of $\eta$ decreases $\alpha$ until it crosses the zero level,  becomes negative (steep spectrum)
and  steeper at larger inclination angles.
It is worth noting that in our work, we can follow 
the change of the spectral index with corresponding displacements of the cores along the jet (Fig.~\ref{fig:alphaVsEta}-b,c). 

\begin{acknowledgements}
We would like to express our gratitude to the  referee for his valuable and profound comments.
GTC thanks INAF-Osservatorio di Arcetri for support.
\end{acknowledgements}

%
\bibliographystyle{aa} 

\end{document}